\documentclass[prb,aps,preprint]{revtex4}

\begin{document}


\title{Electron-spectroscopic investigation of metal-insulator transition
in Sr$_2$Ru$_{1-x}$Ti$_x$O$_4$ ($x$=0.0--0.6)}

\author{Sugata Ray \cite{titech} and D. D. Sarma \cite{jnc}}

\address{Solid State and Structural Chemistry Unit, Indian Institute
of Science, Bangalore 560~012, INDIA \\}


\author{R. Vijayaraghavan}

\address{School of Science and Humanities,
Vellore Institute of Technology Vellore 632 014, INDIA \\}


\begin{abstract}

We investigate the nature and origin of the metal-insulator
transition in Sr$_2$Ru$_{1-x}$Ti$_x$O$_4$ as a function of
increasing Ti content ($x$). Employing detailed core, valence, and
conduction band studies with x-ray and ultraviolet photoelectron
spectroscopies along with Bremsstrahlung isochromat spectroscopy, it
is shown that a hard gap opens up for Ti content $\ge$0.2, while
compositions with $x<$0.2 exhibit finite intensity at the Fermi
energy. This establishes that the metal-insulator transition in this
homovalent substituted series of compounds is driven by Coulomb
interaction leading to the formation of a Mott gap, in contrast to
transitions driven by disorder effects or band filling.

\end{abstract}

\maketitle

PACS number(s): 71.30.+h, 79.60.-i, 71.20.-b, 71.28.+d

\vspace*{1cm}

\newpage

\section{Introduction}

Oxide systems possessing K$_2$NiF$_4$ and related structures have
been extensively studied owing to their interesting structural,
magnetic, and electrical properties. Recently, it was discovered
that Sr$_2$RuO$_4$, a noncuprate member of $A_2B$O$_4$ family,
exhibits~\cite{maeno1} superconductivity at 0.93 K. This report has
naturally provoked many researchers to study the similarity and the
dissimilarity between the superconductivity in this compound and
that in the well-known La$_2$CuO$_4$ with the same crystal
structure, but a much higher $T_c$ [Ref. 2]. The technique of
substituting Cu by impurity ions in La$_2$CuO$_4$ has been
successfully used as a powerful probe to reveal the unconventional
properties of this cuprate.~\cite{maeno2} Similarly, various
attempts have been made with Sr$_2$RuO$_4$ by replacing Ru by Ir
[Ref. 4], Fe [Ref. 5] and Ti [Ref. 6] ions, in order to investigate
the effect of cationic substitutions. Nonmagnetic impurities are
known to induce local moments and suppress $T_c$ severely~\cite{zn}
and therefore, replacement of Ru$^{4+}$ (4$d^4$) ions by nonmagnetic
Ti$^{4+}$ (3$d^0$) ions is an interesting topic of study. The
synthesis of these doped ruthenates is relatively easy, since the
Ti$^{4+}$ ion has the same oxidation state, coordination number as
well as very similar ionic radius when compared with Ru$^{4+}$. In
fact, the compound Sr$_2$TiO$_4$ is iso-structural with
Sr$_2$RuO$_4$ with very similar lattice parameters, making the
substitution facile. It has been reported~\cite{maeno3} that the
Ti$^{4+}$ impurity indeed suppresses the superconductivity and
induces localized moments even for very low Ti concentration, $x$
$<$0.03 in Sr$_2$Ru$_{1-x}$Ti$_x$O$_4$. During the past few years, a
few reports have appeared~\cite{cava1,maeno4,maeno5} investigating
Ti doped Sr$_2$RuO$_4$ system; all these studies focus on the limit
of very low Ti doping. Interestingly, Sr$_2$Ru$_{1-x}$Ti$_x$O$_4$ is
also expected to show a metal to insulator transition as a function
of Ti doping, because the two end members possess widely different
electrical properties with Sr$_2$RuO$_4$ being a paramagnetic metal
and Sr$_2$TiO$_4$ being a band insulator. However, there are only
few reports on the full series (0 $\le x \le$ 1, in
Sr$_2$Ru$_{1-x}$Ti$_x$O$_4$) of compounds~\cite{oswald,vijay}
investigating the transport properties, while no detailed electronic
structure study has been reported to date. These transport studies
reveal that there is a strong anisotropy between the interlayer
resistivity ($\rho_c$) and in-plane resistivity ($\rho_{ab}$),
observed in single crystal Sr$_2$RuO$_4$.~\cite{maeno3,cava1}
$\rho_c$ of Sr$_2$RuO$_4$ shows pure metallic conductivity below 130
K and nonmetallic $d\rho_c$~/~$dT$~$<$~0 at higher temperatures.
However, the metal-nonmetal crossover temperature decreases with
increasing $x$ (Ti concentration) and for $x$=0.2, $\rho_c$ never
enters a metallic regime. The in-plane resistivity remains metallic
throughout the temperature range for Sr$_2$RuO$_4$, while with
increasing $x$, an insulating upturn is observed in the resistivity
data. For $x$=0.2 sample, such semiconducting-like behavior is found
below 40 K [Ref. 9]. Moreover, Oswald {\it et al.}~\cite{oswald}
reported a steep increase in the room temperature resistivity of
polycrystalline Sr$_2$Ru$_{1-x}$Ti$_x$O$_4$ compounds at around $x$
= 0.3. Therefore, all these observations indicate the occurrence of
a metal to insulator transition (MIT) as a function of Ti doping in
this ruthenate series, the critical composition being close to
$x$=0.2.

In order to understand the electronic structure and thereby the
nature of the MIT in this series, we have carried out a detailed
high-energy spectroscopic study on Sr$_2$Ru$_{1-x}$Ti$_x$O$_4$ for
$x$=0.0, 0.1, 0.2, 0.5 and 0.6, employing x-ray photoelectron (XP),
ultraviolet photoelectron (UP), and Bremsstrahlung isochromat (BI)
spectroscopies (S). These experimental studies, supplemented by band
structure calculations for Sr$_2$RuO$_4$ and Sr$_2$TiO$_4$, using an
{\it ab initio} self-consistent linearized-muffin-tin orbital (LMTO)
method within the atomic sphere approximation (ASA), establish the
importance of Coulomb correlation effects in understanding metal
insulator transition as a function of Ti concentration, which can
only be analyzed in terms of Mott transition.

\section{Experimental and theoretical details}

All electron spectroscopic studies reported here were carried out on
polycrystalline samples which were synthesized by the conventional
ceramic route. Stoichiometric quantities of high purity SrCO$_3$ and
the respective transitional metal oxides were thoroughly ground and
the mixture was heated at 600~$^{\circ}$C for 24 h to avoid Ru
evaporation, then at 930~$^{\circ}$C for 24 h, then at
1050~$^{\circ}$C for 24 h, and finally at 1200~$^{\circ}$C for 36 h
with intermittent grindings. The product was then pressed into
pellets and finally sintered at 1200~$^{\circ}$C for 36 hours and
cooled to room temperature in air. The phase purity of each
composition was checked by powder x-ray diffraction (XRD) technique.
A transport study of these compounds has already been
published.~\cite{vijay}

Electron spectroscopic studies were carried out in a commercial
spectrometer from VSW Scientific Instruments Ltd., United Kingdom,
equipped with a monochromatized Al $K\alpha$ photon source for XPS,
a He discharge lamp for UPS, and an electron gun for BIS. The
samples were mounted on copper stubs using ultrahigh vacuum (UHV)
compatible resins, while the electrical contact of the sample with
the instrument ground was maintained using UHV compatible Ag paste.
The pressure of the experimental chamber was better than
5~$\times$~10$^{-10}$ mbar during the experiment and the sample
surface was cleaned {\it in situ} by mechanical scraping using an
alumina file until the various core levels (O 1$s$ and Sr 3$d$)
showed reproducible spectral features and also the intensity of
adventitious C 1$s$ peak was reduced to a minimum. All experiments
were performed at room temperature. The binding energy of all
spectral features were calibrated to the instrument Fermi level
which was determined by recording the Fermi edge region from a clean
silver sample.

Self-consistent LMTO-ASA calculations were performed for
stoichiometric Sr$_2$RuO$_4$ and Sr$_2$TiO$_4$ in the experimentally
determined tetragonal structure with seven atoms per unit cell and
with 128 $k$ points in the irreducible part of the Brillouin zone.
The lattice parameters used for Sr$_2$RuO$_4$ are $a$=$b$=3.8603
\AA~and $c$=12.729 \AA~\ (Ref. 13) and for Sr$_2$TiO$_4$ are $a$=$b$
= 3.88 \AA~and $c$=12.6 \AA~(Ref. 14).
\section{Results and discussion}

In Fig. 1(a), we show the spectral region for Ru 3$p$ and Ti 2$p$
core levels for these compounds. For Sr$_2$RuO$_4$ (\emph{i.e},
$x$=0 sample), we observe a doublet feature with peaks at 463.8 eV
for Ru 3$p_{3/2}$ and 486.4 eV for Ru 3$p_{1/2}$ core level spectra
with a spin-orbit splitting of 22.6 eV. Weak satellite features are
also seen at around 476.5 and 499 eV, corresponding to main Ru
3$p_{3/2}$ and 3$p_{1/2}$ features, respectively. A new distinct
peak appears next to the Ru 3$p_{3/2}$ peak at 458 eV, gaining
intensity monotonically with increasing Ti doping in
Sr$_2$Ru$_{1-x}$Ti$_x$O$_4$; this is identified as the Ti 2$p_{3/2}$
signal. The spin-orbit splitting in 2$p$ level of TiO$_2$~[Ref.15]
is around 5.7 eV, therefore, Ti 2$p_{1/2}$ signal appears at around
463.7 eV binding energy, overlapping the Ru 3$p_{3/2}$ signal. In
order to investigate the Ti 2$p$ core level spectral features
without the overlapping Ru 3$p$ core level signal, we used the Ru
3$p$ signal from the Sr$_2$RuO$_4$ ($x$=0) sample as the reference
spectrum for Ru 3$p$ spectral shape for all the compounds. The basic
idea is to remove the Ru 3$p$ contribution from the spectral region
in Fig. 1(a) by subtracting the reference Ru 3$p$ spectrum, obtained
from the parent undoped Sr$_2$RuO$_4$, after a proper multiplicative
normalization; thus extracted Ti 2$p$ spectra are shown for all
$x$$>$0 in Fig. 1(b). Resulting Ti 2$p_{3/2}$ and 2$p_{1/2}$ peaks
appear at 458 and 463.7 eV binding energies, with the expected
spin-orbit splitting of 5.7 eV for a Ti$^{4+}$ species. Moreover it
is well known that Ti$^{4+}$ oxides, as in SrTiO$_3$~[Ref. 16] or
TiO$_2$~[Refs. 15, 16], show weak satellite features, at about 13 eV
away from the main peak in the Ti 2$p$ spectral region,
characteristic of the Ti$^{4+}$ state. In the present case, we can
see similar weak satellite features at about 13 eV above the main
peaks in the extracted Ti 2$p$ spectra, [marked by arrows in Fig.
1(b)]. This confirms that Ti is in the formally Ti$^{4+}$ 3$d^{0}$
state in this family of compounds, establishing this as a homovalent
substituted (Ti$^{4+}$ in place of Ru$^{4+}$) family of compounds.
It is important to note here that such homovalent substitutions
leave the electronic configuration of Ru sites essentially
unchanged, in contrast to more usual hetrovalent substitutions
(e.g., Sr$^{2+}$ in place of La$^{3+}$) leading to a doping of
charge carriers into the system.

In Fig. 2, we show the calculated density of states (DOS) of the two
end members, namely, Sr$_2$RuO$_4$ and Sr$_2$TiO$_4$. In the upper
panel [Fig. 2(a)], we show the total DOS along with the Ru $d$, O
$p$, Sr $d$, and Sr $s$ partial DOS for Sr$_2$RuO$_4$. The total DOS
has a finite value at $E_F$, consistent with the metallic property
of Sr$_2$RuO$_4$. The occupied part of the DOS is dominated by O $p$
and Ru $d$ states. Significant amounts of O $p$ states extend down
to about -7.8 eV with three distinct groups of features, namely,
those between -3.5 and -7.8 eV, between -1.7 and -3 eV and very
close to the Fermi energy. The Ru $d$ states are found to spread
over the complete valence band range down to -7.8 eV. The feature
near -6 eV is contributed by the bonding O $p$ states with Ru $d$
states. The most intense feature around -2.5 eV in the total DOS
corresponds to primarily O 2$p$ states nonbonding with respect to
Ru-O interactions and therefore, with a negligible contribution from
Ru states. Ru ions in Sr$_2$RuO$_4$ remains in the low-spin
configuration (4$d^4$ : $t_{2g}^4$) because of a strong crystal
field effect~\cite{nakatsuji} splitting the $e_g$ and the $t_{2g}$
levels significantly. Therefore, it has been shown that the states
at $E_F$ arises from the antibonding Ru $d{\epsilon}$($xy, yz, zx$)
and O $p{\pi}$ states having four electron occupancy in these triply
degenerate bands.~\cite{oguchi,dd} The unoccupied DOS between 0 and
3.5 eV is dominated by Ru $d$ $e_g$-derived states, while the states
around 7 eV is contributed mainly by Sr $d$ related states. All the
features above 10 eV in the unoccupied DOS are mainly contributed by
Sr $s$ and $p$ states along with O $s$ states.

The calculated band structure for Sr$_2$TiO$_4$ is shown in Fig.
2(b). It exhibits a clear gap between the occupied and the
unoccupied states, in agreement with the observed insulating
property of the compound, in contrast to the metallic state of
Sr$_2$RuO$_4$. The Fermi level in Fig. 2(b) has been aligned at the
bottom of the conduction band. The calculation indicates a band gap
of 1.3 eV. Expectedly, the occupied part of the band is completely
dominated by the O $p$ states with small contributions from Sr $d$
and Ti $d$ states, consistent with the idea of a formal Ti$^{4+}$
($d^0$) valence state in the compound. The Ti $d$ level has some
occupancy arising from the hybridization of Ti $d$ states with the O
$p$ states. The feature centered at about -5 eV in the total DOS is
found to have significant contributions from such hybridized Ti $d$
states along with the dominant O $p$ contributions. The features
around -3 eV are primarily from an admixture of O $p$ and Sr $d$
states. The first part (0 to 2.7 eV) of the unoccupied density of
states is completely dominated by Ti $d$ states of $t_{2g}$
symmetry, while the Ti $e_g$ states appear between 2.5 and 7.5 eV,
strongly admixed with a contribution from Sr $d$ states. The
features above 12 eV has contributions from Sr $s$, O $s$, and Ti
$p$ states.

The experimental XPS valence band spectra from all the samples are
shown in Fig. 3. The overall spectral features of the valence band
are similar for all the compounds, though there are some systematic
variations in the relative intensities as well as in spectral
details of different features. The XP spectra exhibit two broad
features with peaks at about 1.5 and 5.7 eV binding energies,
respectively. The broad feature centered near 5.7 eV binding energy
is easily seen to be due to primarily O $p$ bonding states for all
compositions; this is evident from the fact that both Sr$_2$RuO$_4$
and Sr$_2$TiO$_4$ have O $p$ DOS over this energy range (see Fig.
2). The spectral intensity of the 1.5 eV feature decreases
monotonically with increasing Ti doping. An intensity decrease of
certain parts over the occupied states is indeed to be expected in
view of a progressive substitution of Ru$^{4+}$ 4$d^4$ state by
Ti$^{4+}$ 3$d^0$ state with an increasing $x$ in this series. A
comparison with the band structure results of Sr$_2$RuO$_4$ [see
Fig. 2(a)] reveals that the spectral feature with the peak at ~1.5
eV is mainly contributed by O $p$ nonbonding states over the higher
energy between 1.5 and 3.5 eV binding energy and the antibonding Ru
$d{\epsilon}$-O $p{\pi}$ states at a lower energy near the Fermi
level. In the case of Sr$_2$TiO$_4$, on the other hand, a gap opens
up near $E_F$ over the energy window where Ru 4$d$ states are found
in Sr$_2$RuO$_4$, due to the 3$d^0$ configuration of the Ti$^{4+}$
in contrast to 4$d^4$ configuration of Ru$^{4+}$. The O $p$
nonbonding contributions in Sr$_2$TiO$_4$ appear around 1.5--2 eV
binding energy as also in the case of Sr$_2$RuO$_4$. Therefore, with
increasing Ti concentration, the Ru-O antibonding states close to
Fermi energy are progressively removed, resulting in a decrease in
the overall spectral intensity in the 0--2 eV binding energy region,
as observed in the experimental spectra. While such considerations
provide an understanding of the overall decrease in the spectral
intensity within the first 2 eV of the Fermi energy with increasing
$x$, these fail to explain a more interesting, though subtle, aspect
of the spectral change. First we note that the spectral intensity at
$E_F$ is substantial for Sr$_2$RuO$_4$, as is evident in Fig. 3,
consistent with its metallic ground state. It is also evident from
the figure that there is negligible intensity at $E_F$ for $x$=0.6,
also consistent with its insulating ground state, thereby capturing
the change in the electronic structure associated with the
experimentally observed metal-insulator transition. However, it is
clear that the previous discussion indicating a decrease in the
spectral intensity will invariably suggest a finite intensity at the
$E_F$, proportional to the Ru content. In other words, the naive
description of an overall decrease of the Ru 4$d$ associated
spectral intensity and the associated changes in the electronic
structure cannot explain the absence of any spectral intensity at
$E_F$ for any finite $x$ less than 1 and therefore, such an approach
cannot explain the metal-insulator transition at a fractional $x$
value in this series. We note that the spectral behavior in the near
$E_F$ region suggests the formation of a gap even at a finite Ru
concentration; such a gap formation is well known in the context of
insulating transition metal compounds, driven by Coulomb
interactions. However, a poor instrumental resolution in XPS
precludes any careful investigation of the gap formation as a
function of $x$ in these spectra. This issue will be discussed in
greater detail at a later stage, making use of the higher resolution
in UP spectra. However, we note here that the correlation effect in
this Ru 4$d$ band system has been shown~\cite{dd} to be unexpectedly
strong and cannot be ignored.

Changes in the electronic structure near the Fermi edge region are
most directly responsible for the metal-insulator transition in any
system and therefore, this spectral region is the most important one
to understand the nature of the MIT in this series of compounds. We
have carefully studied the narrow spectral region close to the Fermi
energy for all the samples, using He {\scriptsize I} UPS, in view of
its much better energy resolution. The set of He {\scriptsize I}
narrow scan spectra, emphasizing the changes in the electronic
structure close to the Fermi energy, is shown in Fig. 4. All the
spectra are normalized at 2.8 eV binding energy, because this
intensity is almost completely derived from O 2$p$ nonbonding states
which remain invariant with Ti doping. We observe nearly the same
intensity for all the spectra at about 2 eV binding energy; these
spectral features are primarily contributed by nonbonding O $p$
states (see Fig. 2) providing an explanation for the near constancy
of this spectral feature across the series. The regular decrease in
spectral intensity with increasing Ti concentration in the 0 to 2 eV
energy region is evident from the figure; this is essentially due to
the substitution of the Ru$^{4+}$ $d^4$ with Ti$^{4+}$ $d^0$
configuration with increasing $x$, decreasing the total electron
count. But more interestingly, we observe a very substantial gap
($\sim$0.1--0.2 eV) opening at $E_F$ for $x\ge$0.2 compounds,
thereby clearly signalling the metal-insulator transition,
consistent with reports~\cite{oswald,vijay} based on transport
measurements.

We have also probed the evolution of the unoccupied states as a
function of Ti doping using BI spectroscopy. The experimental BI
spectra, shown in Fig. 5, exhibit three broad features with peaks at
1.8, 10, and 17 eV . The spectra clearly exhibit finite DOS at $E_F$
for $x$=0 and 0.1 compounds; interestingly, the spectral intensity
at $E_F$ vanishes for $x\ge$0.2, thereby defining an energy gap and
an insulating state, consistent with the valence band results,
discussed above. The feature extending from $E_F$ up to about 5 eV
decreases in intensity systematically with increasing $x$. This can
be understood by considering the cross section of different states.
It is known~\cite{yeh} that the cross section of Ru 4$d$ states is
almost 100 times that of Ti 3$d$ states at an excitation energy of
1486.6 eV, employed in the present study. Therefore, spectra in the
0--3 eV range are dominated by Ru 4$d$ states, with only minor
contributions from Ti 3$d$ states; this immediately explains the
experimental observation of decreasing intensity of the leading
spectral feature with increasing Ti doping in place of Ru. In
contrast to the clear and systematic change in the spectral feature
close to $E_F$, the higher energy features remain almost identical
with changing Ti concentration. This is not surprising, as the band
structure results indeed show that the features above 6 eV are
dominated by O and Sr states, which are essentially the same for
both Sr$_2$RuO$_4$ and Sr$_2$TiO$_4$.

While all these spectroscopic results, establishing a
metal-insulator transition, are consistent with transport
properties,~\cite{oswald,vijay} these spectroscopic results provide
some critical inputs in understanding the possible mechanism
responsible for this MIT, clearly making several other alternative
mechanisms untenable. Though the mechanism of the MIT in this system
has hardly been discussed in the literature so far, our results, for
example, clearly suggest that the MIT in this system cannot be
understood in terms of percolation or any other inhomogeneous model,
with the assumption that the doping of Ti in place of Ru in
Sr$_2$RuO$_4$ does not result in homogeneously doped samples, but
only some inhomogeneous physical mixtures of the end compositions,
namely, Sr$_2$RuO$_4$ and Sr$_2$TiO$_4$. According to such a
description, increasing Ti concentration should gradually increase
the concentration of Sr$_2$TiO$_4$ at the cost of metallic
Sr$_2$RuO$_4$ and consequently, a gradual reduction in the intensity
of the spectral feature at $E_F$ with increasing Ti doping, is
expected. More importantly, the spectra for even the insulating
range of compounds with $x\ge$0.2 are expected to exhibit finite
spectral weight at $E_F$, contributed by the large metallic fraction
present in the inhomogeneous system representing a mixture of
metallic and insulating phases. Evidently, this is not observed
experimentally where a finite and substantial gap is found to appear
in the spectral weight near $E_F$ with only 20\% Ti doping.
Specifically, the changes in the lattice parameters between the end
members [for Sr$_2$RuO$_4$ $a$=$b$=3.8603 \AA~and $c$=12.729 \AA~\
(Ref. 13) and for Sr$_2$TiO$_4$ $a$=$b$=3.88 \AA~and $c$=12.6
\AA~(Ref. 14)] are not large enough to make any perceptible change
in the various hopping strengths that contribute to the bandwidth.
Therefore, the only significant change in the bandwidth will be due
to the decrease in the average coordination number by 20\% at the
same level of doping. Since the bandwidth is linearly proportional
to the coordination number within a tight binding model, we may
expect a decrease of about 20\% in the Ru 4$d$ bandwidth from this
homovalent substitution.

Another possible mechanism for  the MIT is the disorder driven
Anderson transition arising from random replacement of Ru ions by Ti
in the Sr$_2$RuO$_4$ lattice. It is indeed true that the extent of
disorder is expected to increase with increasing Ti concentration at
least until the point of 50\% doping. Such disorder driven
metal-insulator transition is likely to be signalled by a finite
spectral weight at $E_F$, since the Anderson-type transitions are
characterized by finite, but localized, density of states at $E_F$.
In presence of long range Coulomb interactions such a system may
even open up a soft gap, namely, Coulomb gap.~\cite{esp} Electron
spectroscopic investigations have indeed revealed~\cite{gabi} a host
of interesting spectroscopic features associated with disorder
effects in substituted transition metal compounds. However, the
experimental observation of a substantial hard gap in the present
case is clearly not compatible with ideas concerning a
disorder-driven MIT.

The electronic configuration of Ru$^{4+}$ (4$d^4$) ion in
Sr$_2$RuO$_4$ is low-spin. While a strong crystal field would split
the $d$ levels into $t_{2g}$ and $e_g$ states, the Jahn-Teller
distortion in Sr$_2$RuO$_4$ with an elongation of the RuO$_6$
octahedra along the $z$-axis further splits the triply degenerate
$t_{2g}$ level, leading to a stabilization of the doubly degenerate
$d_{zx}$, $d_{yz}$ bands with respect to the $d_{xy}$ band. If this
effect is very strong, the system could go into an insulating state
with a fully filled doubly degenerate $d_{zx}$, $d_{yz}$ band.
However the extent of distortion is not very severe in Sr$_2$RuO$_4$
and the spread of the $d_{xy}$ band is wide enough to overlap with
the $d_{yz, zx}$ band.~\cite{nakatsuji} As a consequence, the system
remains metallic in spite of the modest Jahn-Teller distortion.
Doping with Ti would certainly narrow the Ru bandwidth; however,
this effect is not expected to be large enough to remove the overlap
of the $d_{xy}$ band with the $d_{yz}$ and $d_{zx}$ bands altogether
for as little as 20\% Ti doping.

It has been already reported~\cite{dd} that the electron correlation
effect is not negligible in Sr$_2$RuO$_4$ in spite of the broad Ru
4$d$ band. Therefore, the physical properties of these compounds are
controlled by the relative strengths of on-site Coulomb interaction
energy $U$ and the bandwidth $W$. In the limit of $U$ being
significantly less than $W$, the system is metallic, while a
progressive reduction in $W$ can drive a metal to insulator
transition in a system with integral $d$ occupancy. We believe that
this system is another example of such a bandwidth controlled MIT,
in contrast to any band-filling controlled physics in spite of its
obvious composition-driven change. Our core level study has shown
that Ti retains its stable 4+ oxidation state throughout the series;
as a consequence, the Ru ions throughout the series has an integral
filling with 4$d^4$ electronic configuration. Random replacement of
Ru ions by Ti ions has two distinct effects on the electron motion.
Effects of disorder in such a case will definitely be present;
however, as discussed above, it cannot explain the MIT. The second,
and more important, consequence of the substitution in the present
context is to reduce the Ru-Ru coordination; this evidently leads to
decrease in the Ru 4$d$ bandwidth. Such a progressive reduction in
$W$ with increasing Ti concentration may lead to the value of $U/W$
to cross the critical value for the metal-insulator transition
leading to the insulating state with a finite Mott-Hubbard gap at
the Fermi energy. This description of the MIT appears to be the only
one compatible with all the experimental observations presented
here.

In conclusion, we have investigated the electronic structure of
Sr$_2$Ru$_{1-x}$Ti$_x$O$_4$ with $x$=0.0, 0.1, 0.2, 0.5, and 0.6
using x-ray and ultraviolet photoemission spectra and high-energy
Bremsstrahlung isochromat spectra in conjunction with \emph{ab}
\emph{initio} band structure results. Core level spectra establish
that substituted Ti remains in the Ti$^{4+}$ state throughout the
series, ensuring a Ru$^{4+}$ 4$d^4$ with integral Ru 4$d$ occupancy
per Ru site. Spectra from the occupied and unoccupied parts show the
formation of a gap ($\sim$200 meV) at the Fermi energy for samples
with Ti content, $x$~$\geq$~0.2, establishing a metal-insulator
transition. These available results can only be understood in terms
of a bandwidth controlled metal-insulator transition as a function
of the composition, arising from a decreasing Ru-Ru coordination
with increasing Ti content.

\section{Acknowldgements}

The authors acknowledge the Department of Science and Technology,
and the Board of Research in Nuclear Sciences, Government of India,
for financial support.

\pagebreak

\begin{center}
{\bf FIGURE CAPTIONS}
\end{center}

\vspace*{1.0cm}

\noindent FIG. 1: (a) Ru 3$p$ spectral region with coincident Ti
2$p$ contribution from the Sr$_2$Ru$_{1-x}$Ti$_x$O$_4$ series. (b)
Ti 2$p$ spectra, extracted from the region shown in (a), from the
same.

\vspace*{0.5cm}

\noindent FIG. 2: Calculated density of states (DOS) from
Sr$_2$RuO$_4$ (a) and Sr$_2$TiO$_4$ (b).

\vspace*{0.5cm}

\noindent FIG. 3: XPS valence band spectra from the
Sr$_2$Ru$_{1-x}$Ti$_x$O$_4$ series at 300 K.

\vspace*{0.5cm}

\noindent FIG. 4: He {\scriptsize I} narrow scan UPS spectra from
the Sr$_2$Ru$_{1-x}$Ti$_x$O$_4$ series at 300 K.

\vspace*{0.5cm}

\noindent FIG. 5: BIS spectra from the Sr$_2$Ru$_{1-x}$Ti$_x$O$_4$
series at 300 K.

\end{document}